\documentclass[a4paper,10pt]{article}

\usepackage{psfrag}
\usepackage{chemarr}
\usepackage{amssymb}
\usepackage{multirow}
\usepackage{amsmath}
\usepackage{xr}
\usepackage{psfrag}
\usepackage{array}
\usepackage{natbib}
\usepackage{epstopdf}
\usepackage{appendix}

\usepackage{graphicx}

\newcommand{\supportingsection}[2][]{\setcounter{figure}{0}\renewcommand{\thefigure}{S\arabic{figure}}\setcounter{table}{0}\renewcommand{\thetable}{S\arabic{table}}}

\newcommand{\ecoli}{{\em E.coli}}
\newcommand{\one}{($i$) }
\newcommand{\two}{($ii$) }
\newcommand{\three}{($iii$) }
\newcommand{\four}{($iv$) }
\newcommand{\five}{($v$) }

\newcommand{\onel}{($a$) }
\newcommand{\twol}{($b$) }

\usepackage[dvips,margin=2cm]{geometry}

\date{}

\begin{document} 
\title{The effects of transcription factor competition on gene regulation.} 
\author{Nicolae Radu Zabet$^{1,2,\ast}$ and Boris Adryan$^{1,2,\dagger}$    \\
\mbox{}\\
\footnotesize $^1$Cambridge Systems Biology Centre, University of Cambridge, Tennis Court Road, Cambridge CB2 1QR, UK\\
\footnotesize $^2$Department of Genetics, University of Cambridge, Downing Street, Cambridge CB2 3EH, UK\\
\footnotesize $^\ast$Email: n.r.zabet@gen.cam.ac.uk\ \ \footnotesize $^\dagger$Email: ba255@cam.ac.uk}

\maketitle


\begin{abstract}

Transcription factor  (TF) molecules translocate by \emph{facilitated diffusion} (a combination of 3D diffusion around and 1D random walk on the DNA). Despite the attention this mechanism received in the last $40$ years, only a few studies investigated the influence of the cellular environment on the facilitated diffusion mechanism and, in particular, the influence of `other' DNA binding proteins competing with the TF molecules for DNA space. Molecular crowding on the DNA is likely to influence the association rate of TFs to their target site and the steady state occupancy of those sites, but it is still not clear how it influences the search in a genome-wide context, when the model includes biologically relevant parameters (such as: TF abundance, TF affinity for DNA and TF dynamics on the DNA).

We performed stochastic simulations of TFs performing the \emph{facilitated diffusion} mechanism, and considered various abundances of cognate and non-cognate TFs. We show that, for both obstacles that move on the DNA and obstacles that are fixed on the DNA, changes in search time are not statistically significant in case of biologically relevant crowding levels on the DNA. In the case of non-cognate proteins that slide on the DNA, molecular crowding on the DNA always leads to statistically significant lower levels of occupancy, which may confer a general mechanism to control gene activity levels globally. When the `other' molecules are immobile on the DNA, we found a completely different behaviour, namely: the occupancy of the target site is always increased by higher molecular crowding on the DNA. Finally, we show that crowding on the DNA may increase transcriptional noise through increased variability of the occupancy time of the target sites.

For biologically relevant crowding levels, molecular crowding on the DNA does not significantly influence the association rate of TFs to their target site (independent of whether the `other' molecules are mobile or fixed on the DNA), but it significantly affects the occupancy of the target sites and the associated noise (for both fixed and immobile obstacles on the DNA). 
\end{abstract}


\section{Introduction}
Transcription factors (TF) are DNA-binding proteins that regulate gene activity by binding to specific sites on the DNA. \citet{riggs_1970a} observed that the association rate of the lac repressor (a bacterial TF) to its target site is much faster than predicted by simple 3D diffusion. It was later proposed that the mechanism by which TF molecules locate their target sites assumes a combination of 3D diffusion and 1D random walk on the DNA, which is often called \emph{facilitated diffusion} \citep{berg_1981,halford_2004}. Their rationale was that the speed-up in target site finding is achieved by reducing the dimensionality of the search process. The existence of facilitated diffusion was proven experimentally both \emph{in vitro} \citep{kabata_1993} and \emph{in vivo} \citep{elf_2007}. 

Following this initial work, a large number of theoretical studies investigated the search process and described the effects that various factors have on the speed at which TFs locate their target sites. With a few exceptions, these studies considered the case of one TF molecule performing the search process on naked DNA, without any competitor species. It is clear that this is an approximation that needs further investigation, because other proteins, including TFs with different specificity, are translocating on the DNA at the same time. In fact, the proportion of inaccessible DNA is high; for example, between $10\%$ and $50\%$ of the \emph{E.coli} DNA is bound by other proteins (which we call `non-cognate') \citep{flyvbjerg_2006}. 

Usually, it is assumed that only one molecule performs the random search \citep{halford_2004,mirny_2009}, but in bacterial cells, TFs usually display $10-100$ copies per cell \citep{wunderlich_2009}. Thus, the TF copy number could potentially influence the search time \citep{fofanno_2012} and, consequently, the amount of time the target sites are occupied. 

The question that we address in this manuscript is: how does the abundance of TFs and the presence of other molecules on the DNA influence TF target site finding and binding? In particular, we are interested in describing both the mean and the variability (`noise') of the association rate to a specific target site and of the proportion of time this target site is occupied. 

There is a notion that crowding on the DNA can have two opposing effects: \one reducing the amount of DNA that needs to be `scanned' by covering non-specific sites \citep{mirny_2009} and \two increasing the probability that the target site is already covered by non-cognate molecules \citep{flyvbjerg_2006}. In other words, by increasing the abundance of non-cognate molecules, the amount of DNA that needs to be scanned is reduced, but, at the same time, the probability that the target site is occupied by a non-cognate molecule is increased. This suggests that there may be a level of DNA occupancy which optimises the search speed. 

\citet{murugan_2010} proved the existence of an optimal amount of crowding analytically, but their approach contained approximations that could introduce biases in the final results. One of their assumptions was that the sliding length is inversely proportional to the number of molecules bound to the DNA, which is true only if a bound molecule performs just 1D random walks and does not hop or jump, which are commonly accepted modes of TF translocation \citep{bonnet_2008,wunderlich_2008}. Furthermore, \citet{murugan_2010} disregarded the fact that the non-specific association rate is decreased when the DNA is occupied by other molecules and that the target site can also be occupied by non-cognate molecules. When these aspects are taken into account, \citet{li_2009} showed that the time to locate the target site always increases with increasing amounts of crowding on the DNA. However, aforementioned studies \citep{flyvbjerg_2006,murugan_2010,li_2009} assumed that the proteins bound to the DNA act as fixed obstacles, i.e. they do not move on the DNA. This approximation needs further analysis, because non-cognate TF molecules will display similar dynamic behaviour to the cognate TFs under investigation. 

\citet{marcovitz_2013} addressed the question of the difference between mobile and immobile obstacles and found that, in the case of immobile obstacles, there is a crowding level that minimises the search time, while, in the case of mobile obstacles, the search time grows monotonically with increasing crowding levels. Their model displayed a higher level of detail (by representing explicitly the 3D structure of the DNA and the 3D diffusion of molecules), which meant that they could only focus on a small system of $100\ bp$ of DNA and obstacles covering $2\ bp$. While this model might accurately represent an \emph{in vitro} system, the size of the DNA is prone to affect the applicability of the results for \emph{in vivo} systems (where the model has to consider the entire genome); as we   proposed in \citep{zabet_2012_subsystem}.  

As an important step from these previous studies (that were either restricted to smaller subsystems that are relevant only for \emph{in vitro} studies, or relied on mean field approximations), we address the question of how molecular crowding on the DNA influences the TF search process, and the occupancy of the target site, in the context of a comprehensive model of the facilitated diffusion mechanism \citep{zabet_2012_grip,zabet_2012_model}. In particular, our model considers the entire DNA with multiple DNA binding molecules and is fed with parameters that were estimated from experimental measurements (which leads to biologically relevant representation of the bacterial cells). Using a well-characterised TF and its best known binding site as a model, our results indicate that the average time the \emph{E.coli} lac repressor (lacI) requires to locate the $O_1$ site is increased with the addition of non-cognate molecules that move on the DNA (supporting the result of \citet{li_2009}), while, in the case of fixed roadblocks on the DNA, there seems to be a crowding level that optimises the mean of the search time (supporting the results of \citet{murugan_2010}). Nevertheless, we found that the changes in the arrival times are not statistically significant, in the case of biologically relevant crowding levels (between $10\%$ and $50\%$ of the DNA being covered by DNA binding molecules), for both mobile and immobile obstacles. 

Finally, we also measured the time the $O_1$ site was occupied by a lacI molecule during one hypothetical \emph{E.coli} cell cycle. The results show that, in the case of obstacles moving on the DNA, crowding decreases the average target site occupancy time (and this is statistically significant), while simultaneously the variation in occupancy is significantly increased. This means that  noise can, in part, be accounted by the inherent crowding of molecules on the DNA and is supported by recent experimental evidence that non-cognate TFs contribute to gene expression noise \citep{sasson_2012}. In the case of fixed obstacles, we found the opposite effect, namely that increasing the crowding always leads to a statistically significant increase in the occupancy of the target site, but at the same time it also leads to a higher probability that the target site is never reached within the cell cycle. This suggests that in the case of fixed obstacles on the DNA, higher crowding can lead to a  binary behaviour of the occupancy of the target site (the target sites are occupied in fewer cells, but when they are occupied, they can display significant increase in occupancy time). 

\section{Results}

\subsection{Time to locate the target site.} 

\begin{figure}
\centering
\includegraphics[angle=270,width=\textwidth]{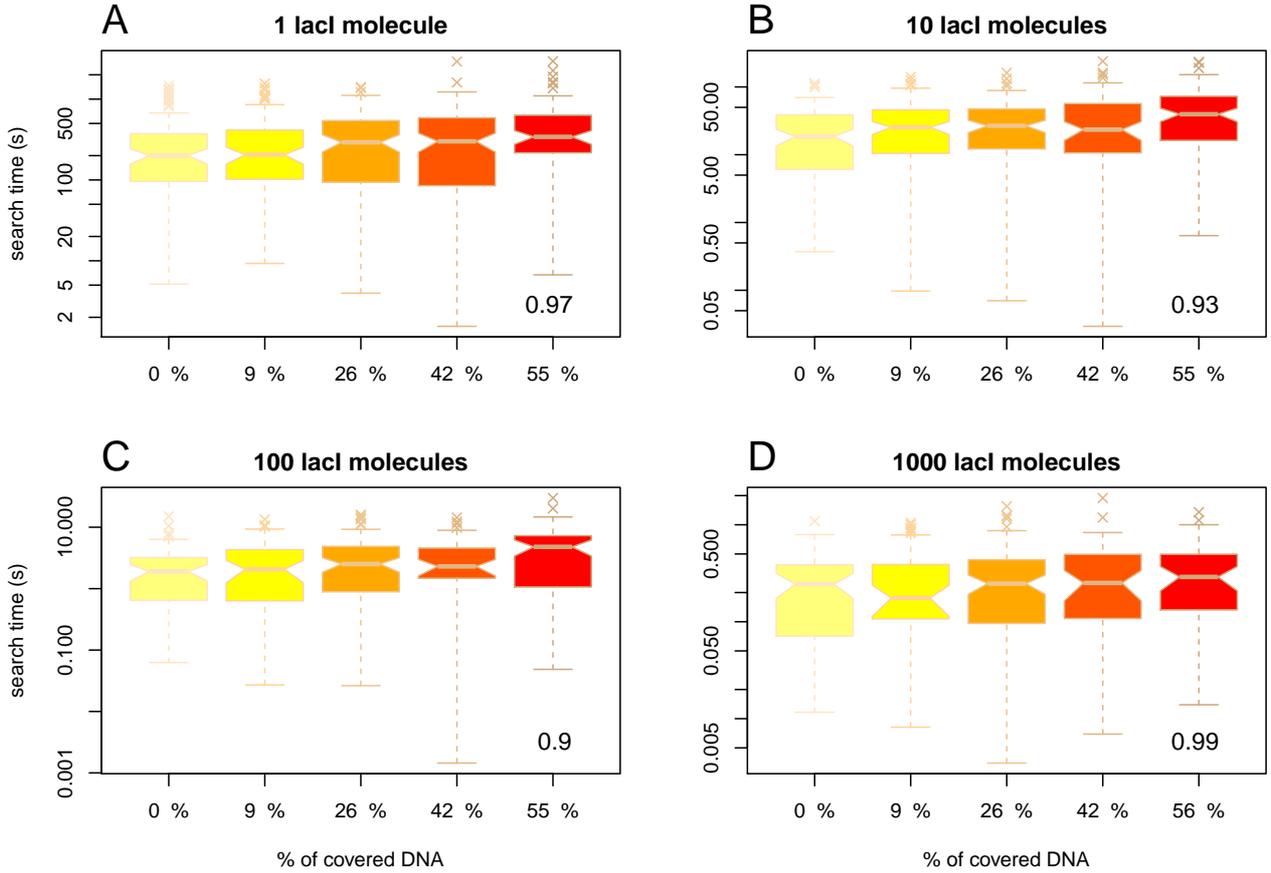}
\caption{\emph{The average time for the TF to reach the target site (measured in seconds) as a function of DNA crowding in the case of mobile obstacles}. Note the differences between scales of the y-axis, e.g. for $1$ lacI molecule it takes in the range of tens of minutes to locate the target site, while for $100$ copies the search time is in the range of seconds. The number in the inset represents the Pearson coefficient of correlation between crowding and the mean of the search time. The values indicate the crowding is highly correlated with the search time, in the sense that higher crowding on the DNA leads to higher search times. Note that to enhance the visibility, the boxplots are positioned equidistant although the crowding levels are not. \label{fig:TStime}}
\end{figure}

First, we wanted to understand how crowding influences the association rate of a TF to its target site. Figure  \ref{fig:TStime} shows the arrival times of the first lacI molecule to the $O_1$ site for various abundances of non-cognate TFs and lacI. Figure  \ref{fig:TStime}\onel\  considers the case of $1$ lacI molecule in the cell and several levels of crowding on the DNA and shows that, by increasing the amount of crowding, the mean arrival times always increase, but there is negligible change in the variance of the search time in the range of biologically relevant crowding levels on the DNA. 

Nevertheless, we found that the increase in the search time is not statistically significant. In particular, we performed a Tukey's range test (for a $95\%$ confidence interval) in conjunction with a one-way ANOVA, which revealed that only in the case of $1$ or $10$ molecules of lacI, and crowding levels of at least $55\%$ on the DNA, there is a statistically significant difference in the search time; see the Figure \ref{fig:TStimeANOVA}. This result suggests that, for crowding within biologically relevant levels (between $10\%$ and $50\%$ of the DNA being covered by DNA binding proteins), the molecular crowding on the DNA has negligible effects. \citet{li_2009}, found similar results, in the sense that they observed a low increase in the search time for crowding levels within biologically plausible levels. Nevertheless, since they performed an analytical study, they were able to identify only the mean search time, while here we show that when variability in the arrival time is included in the analysis, the increase in the search time become negligible.   

Next, we wanted to confirm that the results of our simulations were in accordance with previous experimental studies. For example,  \citet{elf_2007} found that the time of $1$ lacI molecule to locate the $O_1$ site is $\approx 354\ s$. For $10$ molecules of lacI (which is the endogenous level of lacI in \emph{E.coli}) the search time will be ten times faster, $\approx 35\ s$. Figure \ref{fig:TStime}\twol\  shows that in our simulations $10$ lacI molecules can locate the $O_1$ site on average within similar times, but only for a degree of crowding levels of: $9\%$ ($\left\langle T_{}^{0.09}\right\rangle=35.84\ s$), $26\%$ ($\left\langle T_{}^{0.26}\right\rangle=35.52\ s$) and $\left\langle T_{}^{0.42}\right\rangle=38.13\ s$). If there is no competition on the DNA, the time is shorter ($\left\langle T_{}^{0}\right\rangle=27.02\ s$), while for higher levels of crowding the time is higher ($\left\langle T_{}^{0.55}\right\rangle=52.05\ s$). This confirms that the system was correctly parametrized and that for biologically plausible crowding levels we obtain similar results to the experimental measurements. We can conclude that the arrival time for all considered crowding levels deviates only negligibly  from the experimentally measured value; see also Figure \ref{fig:TStimeANOVA}. Furthermore, in the case of empty DNA, the mean search time is similar to the one proposed by \citet{bauer_2013}, when they considered empty DNA and the 3D organisation of the \emph{E.coli} genome. This suggests that the 3D organisation of the \emph{E.coli} genome has only a limited effect on the search time.

One difference between our model and previous models \citep{li_2009,murugan_2010} is that we assumed mobile obstacles, while the previous models assumed immobile obstacles. To investigate the impact of this assumption, we also performed a series of simulations where we considered the non-cognate TFs to be immobile obstacles as in \citep{li_2009}. The description of this `TF species' can be found in the methods section. In the case of immobile obstacles on the DNA, there is a different functional relationship between crowding on the DNA and the amount of time required by a TF to bind to its target site, in the sense that there is a crowding level on the DNA (or an interval of crowding)  that minimises the mean of the search time, thus, supporting the findings of \citet{murugan_2010}; see Figure \ref{fig:TStimeImmobile}. Although visually difficult to notice, we found that, in the case of $40\%$ of the DNA being covered by immobile obstacles, there is a minimum in the mean of the search time. For example, if one lacI molecule needs on average $282\ s$ to locate its target site in the case of naked DNA,  then, in the case of $40\%$ of the DNA being covered by immobile non-cognate molecules, the mean search time reduces to $244\ s$. Increasing the crowding level about this value leads to an increase in the search time up to $417\ s$ (in the case of $70\%$ of the DNA being covered by immobile obstacles). In  \emph{E.coli}, there seem to be $\approx3\times10^4$ molecules on the genomic DNA \citep{murugan_2010}, which potentially suggests that the abundance of DNA binding proteins in \emph{E.coli} is set to minimise the search time of TFs for their target site. Nevertheless, these changes in search time are not statistically significant except for crowding levels of $70\%$, which suggests that, for biologically relevant crowding levels on the DNA (between $10\%$ and $50\%$ \citep{flyvbjerg_2006}), the search time is not significantly affected by the molecular crowding on the DNA  or by the fact that the obstacles are mobile or immobile.

\subsection{Proportion of time the target site is occupied.} 

The second aspect we were interested in is the proportion of time the target site is occupied by cognate TFs, as this may have direct influence on gene expression. \citet{sasson_2012} found that binding sites of genes that are occupied by cognate TF molecules for shorter amounts of time display a larger degree of gene expression noise compared to binding sites that are occupied for longer times. They attributed this noise to the fact that cognate TF molecules can `insulate' the target site from non-cognate TF molecules. We wanted to verify the validity of this assumption and, thus, we measured the fraction of time the target site is occupied during stochastic simulation of the facilitated diffusion mechanism. 

\begin{figure}
\centering
\includegraphics[angle=270,width=\textwidth]{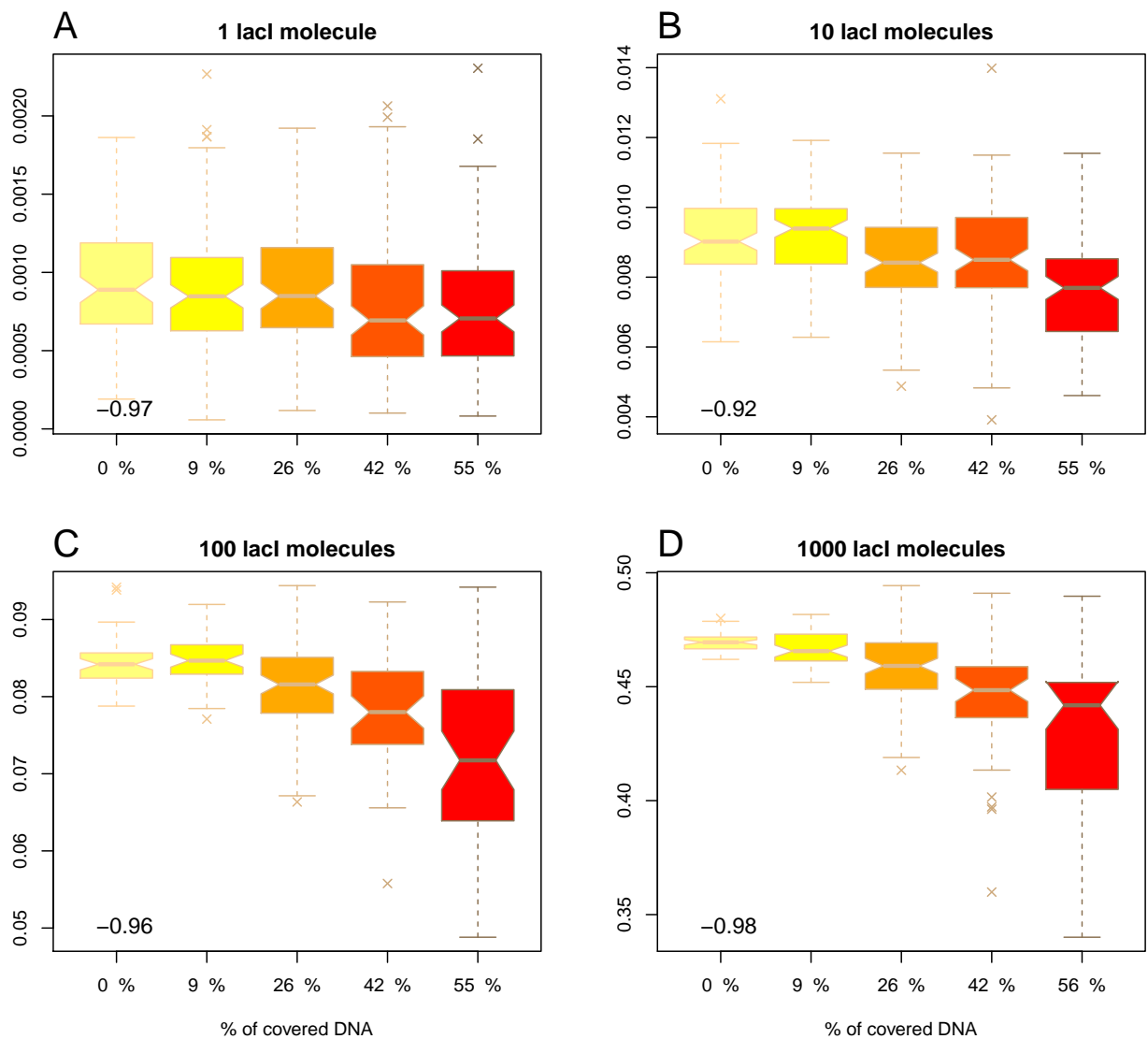}
\caption{\emph{Proportion of time relative to the cell cycle that the target site is occupied (y-axis) as a function of DNA crowding (x-axis)  in the case of mobile obstacles}. The number in the inset represents the Pearson coefficient of correlation between crowding and the mean of the proportion of time the $O_1$ site is occupied. The values indicate that crowding is highly anti-correlated with the proportion of time the target site is occupied, in the sense that higher crowding on the DNA leads to lower occupancy of the target site by cognate TFs.  Note that to enhance the visibility, the boxplots are positioned equidistant although the crowding levels are not. \label{fig:TSproportion}}
\end{figure}

Figure  \ref{fig:TSproportion} shows that molecular crowding on the DNA reduces the average occupancy of the target site, as previously proposed by  \citet{wasson_2009}, and this reduction in occupancy is statistically significant (except in the case of $1$ cognate molecule, which is usually attributed to leaky expression of the gene encoding the TF); see the Figure \ref{fig:TSproportionANOVA}. In the case of 10 molecules of lacI, the occupancy is reduced by $17\%$ when the crowding increases from $9\%$ to $55\%$. This means that crowding on the DNA can control gene expression levels at a global level. In the case of activating TFs, the increase in DNA-binding protein copy numbers may lead to a reduction in gene expression.

Furthermore, this reduction in the average occupancy also introduces a larger degree of variability that can be observed at target sites; see Figure  \ref{fig:TSproportion}. For example, in the case of 10 lacI molecules, the variance almost doubles (increase by $80\%$), when the crowding is increased from $9\%$ to $55\%$. This higher variability, in conjunction with the lower occupancy of the target site, may result in an amplified increase of the noise in gene regulation; see Figure \ref{fig:TSproportionNoise}. One method to reduce the noise levels in the occupancy of the target site is increasing the abundance of the cognate TF (lacI in our case)  \citep{becskei_2005,paulsson_2005,even_2006,zabet_2009}. Our results confirm that the increase in the noise levels generated by crowding can be compensated by an increase in lacI copy number.

\begin{figure}
\centering
\includegraphics[angle=270,width=\textwidth]{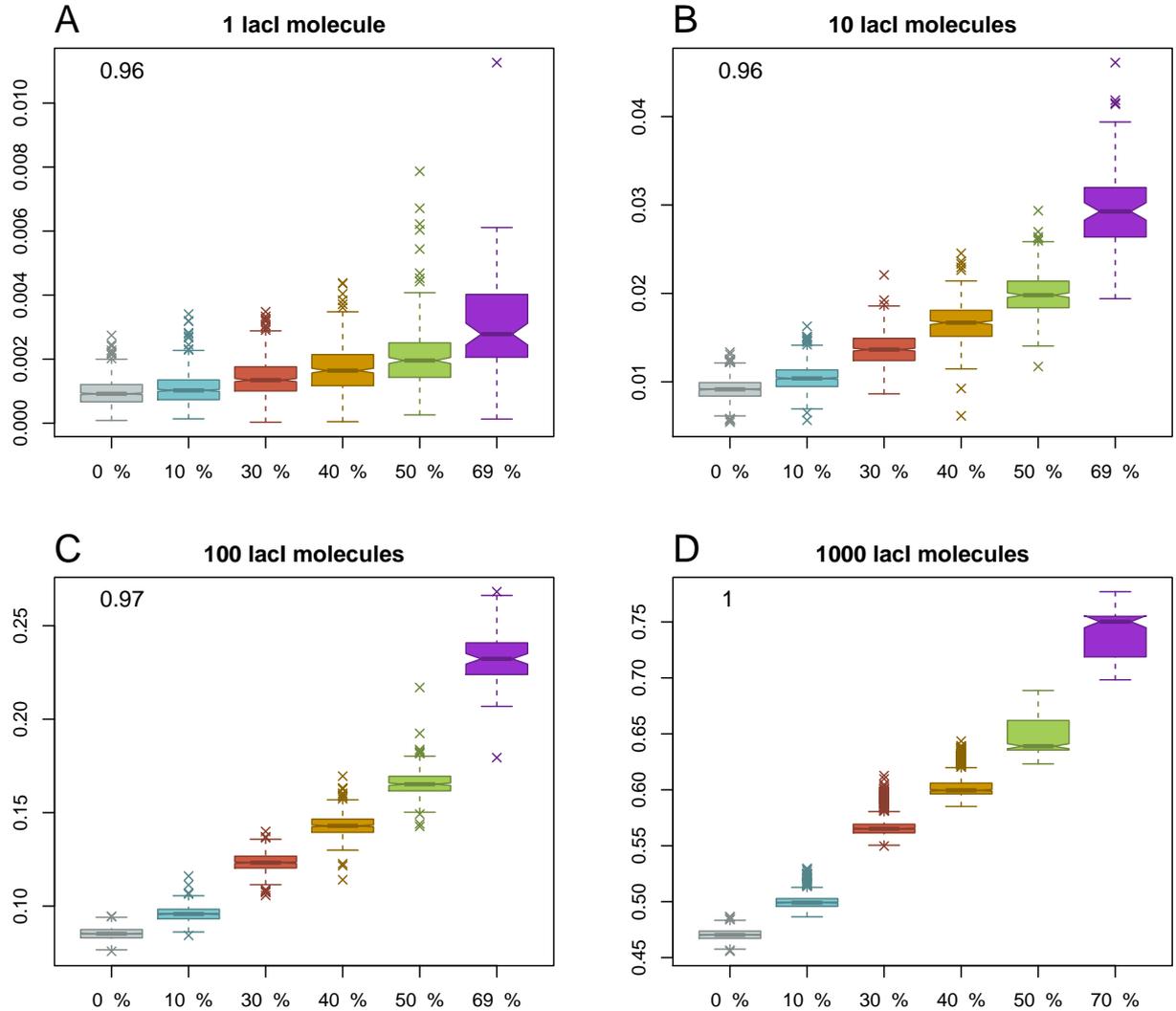}
\caption{\emph{Proportion of time relative to the cell cycle that the target site is occupied (y-axis) as a function of DNA crowding (x-axis)  in the case of immobile obstacles}.For each set of parameters, we performed $1000$ simulations. The mean occupancy of the target site is highly correlated with the crowding level on the DNA for all lacI abundances. Note that for higher crowding the number of simulations where the target site is reached within $3000\ s$ decreases and, for $70\%$ of the DNA being covered by DNA binding proteins, the probability to locate the target site drops to $0.1$; see Figure \ref{fig:TS3Dcount}.  The number in the inset represents the Pearson coefficient of correlation between crowding and the mean of the proportion of time the $O_1$ site is occupied. The values indicate that crowding is highly correlated with the proportion of time the target site is occupied, in the sense that higher crowding on the DNA leads to higher occupancy of the target site by cognate TFs.  Note that to enhance the visibility, the boxplots are positioned equidistant although the crowding levels are not. \label{fig:TSproportionImmobile}}
\end{figure}

Finally, we considered again the case of immobile obstacles and measured the occupancy of the $O_1$ site. Figure \ref{fig:TSproportionImmobile} displays an unexpected effect, namely that by increasing the crowding level, the occupancy of the target site increases as well and, again, this change is statistically significant; see Figure \ref{fig:TSproportionANOVA3D}. The explanation for this result is that by increasing the molecular crowding on the DNA, the cognate molecules are confined more time in the vicinity of the target site as proposed by \citet{wang_2012}. Nevertheless, in conjunction with this increase in occupancy of the target site, there is also a decrease in the number of simulations where the target site is reached.  In other words, by increasing the crowding level on the DNA there are fewer cases where the target site is reached within one cell cycle ($3000\ s$), but when (i.e., if) the target site is reached, the occupancy is higher, suggesting a change from a graded behaviour (in the case of mobile obstacles) to a binary behaviour (in the case of immobile obstacles).

\section{Discussion}
The influence that molecular crowding has on gene regulation has been considered only in a few previous studies. These studies mainly focused on the mean arrival time to the target site (such as \citep{murugan_2010} and \citep{li_2009}) or variability of target site occupancy \citep{sasson_2012}. Although these works provided analytical solutions on this issue, they did not consider the case of `mobile obstacles' on the DNA \citep{zabet_2012_review}. Here, we performed stochastic simulations where each molecule was explicitly represented, thus allowing an assessment of the difference between  ‘mobile’ versus ‘fixed’ obstacles. 

Our results show that, in the case of immobile obstacles on the DNA, there is a crowding level that minimises the mean of the search time (as found by \citet{murugan_2010}), while, in the case of mobile obstacles, molecular crowding on the DNA (implemented through the presence of non-cognate TFs) increases the arrival time of cognate TFs to their target site (as previously proposed by \citet{li_2009}). This increase in search time for high crowding levels on the DNA could potentially be explained by barriers forming in the vicinity of the target site as suggested in \citep{ruusala_1992,hammar_2012,wang_2012}. Nevertheless, we found that, within biologically relevant crowding levels, these changes in search time were small. Recently, \citet{marcovitz_2013} found similar results (for immobile obstacles there is a crowding level that minimises the search time and for mobile obstacles the search time increases monotonically with the crowding levels), when they represented explicitly the 3D organisation of the DNA and the 3D diffusion of the molecules. Note that this work studied the proportion of scanned nucleotides, which is related to the time required to locate the target site. However, they considered  only $100\ bp$ of DNA and obstacles that cover only $2\ bp$, which can introduce biases in the results if we consider real biological systems \citep{zabet_2012_subsystem}. For example, in \ecoli, molecules perform facilitated diffusion on $\approx 4.6\ Mbp$ (genome-wide) \citep{riley_2006} and they cover on average around $20\ bp$ when bound to the DNA \citep{stormo_1998}. In fact, we found that the change in arrival time, introduced by molecular crowding on the DNA, is not statistically significant for biologically plausible crowding levels in bacterial cells (in \emph{E.coli}, between $10\%$ and $50\%$ of the DNA is covered by DNA binding proteins \citep{flyvbjerg_2006}) irrespective of whether the obstacles on the DNA are mobile or fixed.  From this, one can conclude that the TF search time in bacterial cells is robust to changes in the molecular crowding level on the DNA. This result has a twofold implication: \one for biologically relevant crowding levels on the DNA, the search time is not significantly affected by molecular crowding and \two, there is no statistically significant difference between fixed and mobile obstacles on the DNA with respect to the search time for biologically relevant crowding levels. 

Importantly and in contrast to the search time, in the case of mobile obstacles on the DNA, crowding leads to a reduction in the proportion of time the target site is occupied and this reduction in occupancy is statistically significant. This may be an important feedback mechanism in cases where genes encode transcription factors. For example, in the case of activator transcription factors, an increase in activator TFs abundance will lead to an increase in crowding, which, consequently, results in a reduction of the binding of activator TFs to their target sites (thus, resulting in negative feedback). Analogously, in the case of repressing transcription factors, if the repression is achieved by blocking the binding of RNA polymerase to promoters, then an increase in crowding on the DNA would lead to further repression (again, resulting in negative feedback). 

Genetic research and synthetic biology often employ experiments where the abundances of one or several TFs are changed significantly (either completely knocked down or significantly over-expressed). The general assumption is that only the genes that are directly regulated by the corresponding TFs (and to some extent their downstream targets) will be affected by this change. Nevertheless, significant changes in the overall abundance of DNA-binding proteins can lead to changes of the crowding on the DNA. Our study suggests that, in that case, the activity state of all genes can be affected by the changed degree of crowding. It can be assumed that evolution has come up with compensatory mechanisms that guarantee stable genomic expression levels, or that the degree of crowding must change significantly (beyond what is biologically feasible) for these effects to be measurable. 
This is where stochastic simulations can only inform us of theoretical possibilities, but where ultimately biological experiments are required.

In the case of immobile obstacles, crowding leads to a statistically significant increase in the occupancy of the target site, but, at the same time, the proportion of simulations where the target site is reached within a cell cycle drops significantly (mainly due to total or partial blockage of the target site by immobile obstacles). A bioinformatics study performed by \citet{hermsen_2006} revealed that, in \emph{E.coli}, TF binding sites often overlap (they found that $39\%$ of the binding sites overlap at least once) and this indicates that the exclusion of TFs from their target sites by molecular crowding on the DNA is a biologically plausible scenario. These two opposite effects suggest that, in the case of high number of fixed obstacles on the DNA, within the population the occupancy of the target sites display binary response, in the sense that in a subset of `virtual' cells the target site is never reached, but, in the rest of the `virtual' cells, the occupancy of the target site is greatly increased mainly due to the confinement of the TF molecule in the vicinity of the target site \citep{wang_2012}.

In both cases (mobile and immobile obstacles), crowding  causes an increase in variability of the occupancy state across the population. Note that the variability here refers to population level variability and not time fluctuations, i.e., each simulation considers an independent `virtual' cell. This means that a cell that has a lower number of DNA-binding proteins may display a finer control on gene regulation and less gene regulation noise. In order to get more local control on gene regulation, lower crowding on the DNA is required, but crowding on the DNA in unavoidable. Hence, when the cell grows too much (in the sense of overall protein production) and the DNA gets overcrowded, the noise in gene regulation reduces the fitness of the cell, an aspect which can be compensated only if the cognate TF abundance increases as well. This indicates that when the cognate TFs are a fixed percentage of the total abundance of DNA-binding proteins, there is an optimal level of crowding above which the noise in gene regulation becomes harmful for the cell (similar to the results of \citet{li_2009}).

Often it is assumed that there is a direct relationship between binding site occupancy and expression level. We show that the variability in occupancy is not negligible and depends on the number of non-cognate molecules bound to the DNA. This variability that can be observed between cells, is independent of fluctuations in the TF abundances (cognate or non-cognate), but arises from the facilitated diffusion mechanism and depends on crowding. In contrast,  \citet{bauer_2013} found negligible variability in the search time, when they modelled the facilitated diffusion process assuming the 3D organisation of the \emph{E.coli} genome, but discarding the affinity landscapes of the TF. Here we show that the search time displays high variability when considering the TF affinity landscape, but this variability is not influenced significantly by the crowding levels on the DNA (in the case of mobile obstacles on the DNA); see Figure \ref{fig:TStime}. In this context, the omission of variations in occupancy of the \emph{cis-}regulatory region or wrong assumptions about its extent can generate misleading results when investigating the sources of noise in gene expression.

Overall, we found that only for immobile obstacles the occupancy of the target site is significantly higher (while  the search time is is only negligibly affected within biologically relevant levels of molecular crowding on the DNA, for both mobile and immobile obstacles); see Figure \ref{fig:TStimeImmobile}. This shows again how important the underlying assumption of immobile versus mobile obstacles is, in the case of genomic occupancy of TFs.

\citet{slutsky_2004b} identified that the TF target search process is affected by the so-called speed-stability paradox, where the search process can be fast and lead to weak binding to the target site, or the search process can be slow and lead to strong binding to the target site. In the case of immobile obstacles, we showed that high crowding levels (which are within biologically plausible values) lead to higher occupancy at the target site and and at the same the search time is not significantly affected. This suggests that the presence of immobile obstacles can potentially reduce the effects of the speed-stability paradox.

In this context, one might ask whether highly abundant fixed obstacles on the DNA really exist? In bacterial cells, given the high specificity of some TFs, we expect that a subset of the TFs would potentially create these immobile obstacles (e.g. CRP). However, given the low abundance of most other bacterial TFs \citep{wunderlich_2009}, the position where these immobile obstacles emerge is encoded into the DNA. Thus, we cannot make a general statement regarding the molecular crowding on the DNA, but suggest this needs more systematic analysis for each particular promoter region.

Alternatively, barriers can form on the DNA when there is strong direct TF-TF cooperativity, which will lead to cluster formation on the DNA \citep{chu_2009}. This effect is removed when non-cognate TFs (that do not display direct TF-TF cooperativity) are present in the cell, but it is always the case that molecules that do not display cooperativity will be bound to the DNA.  

Finally, the presence of nucleosomes on the DNA could be responsible for these barriers, but this is particular only for eukaryotic systems and there is still no clear evidence in what form facilitated diffusion exists in eukaryotic cells \citep{vukojevic_2010,gehring_2011}; discussed in \citep{zabet_2012_review}. 

\section{Materials and Methods}

We performed stochastic simulations using a computational framework and a set of parameters presented in \citep{zabet_2012_grip,zabet_2012_model}. Briefly, the model represents explicitly all molecules in the system and allows to perform event driven stochastic simulations of the dynamics of the molecules in the system \citep{gillespie_1976,gillespie_1977}. The 3D diffusion is modelled implicitly by using the Master Equation, which was shown to be an accurate approximation when simulating binding of TFs molecules to the DNA \citep{zon_2006}. The molecular crowding in the cytoplasm only scales the binding equilibrium constant and the 3D diffusion constant \citep{morelli_2011}. Our method to estimate the association rate to the DNA  \citep{zabet_2012_model} ensures that the TF molecules are bound to the DNA approximately $90\%$ of the time (as it was experimentally measured in \citep{elf_2007}) and this means that the effects of crowding in cytoplasm are implicitly incorporated in our model (through the association rate to the DNA). 

Furthermore, our model assumes that the DNA is a string of letters \verb+{A,C,G,T}+ and, thus, we disregard the 3D organisation of the \emph{E.coli} genome. This aspect, the 3D organisation of the genome, could potentially influence the search time as shown in \citep{brackley_2012,fofanno_2012}. \citet{bauer_2013} considered a coarse grained model of the 3D structure of the \emph{E.coli} genome and found that in the case of $1$ TF molecule and empty DNA the mean search time is approximately $311\ s$. This value is similar to our result for empty DNA and $1$ molecule of lacI searching on the DNA ($282\ s$) and, thus, it seems that including the 3D organisation of the  \emph{E.coli} genome would lead to only small deviations from our results.

The amount of time a molecule spends at a certain position on the DNA is a random number exponentially distributed with an average which is determined based on the binding energy \citep{gerland_2002}, here, approximated by the position weight matrix \citep{stormo_2000}. Once the amount of time spent at one position expires, the molecule can slide to a nearby position, hop on the DNA or unbind from the DNA with certain probabilities which were previously estimated in \citep{zabet_2012_model}. Finally, steric hindrance is implemented by not allowing two molecules to cover the same base pair simultaneously \citep{hermsen_2006}. In our system, we assume the existence of two TF species: a cognate (lac repressor in our case) and a non-cognate. The parameters associated with the lac repressor, including its specificty expressed as position weight matrix, can be found in Table \ref{tab:modelTFparams} and Table \ref{tab:lacI3gPWM}.

\subsection{System size reduction}

In \citep{zabet_2012_subsystem} we showed that it is sufficient to simulate the target finding process using a smaller (of at least $100\ Kbp$) region of DNA, provided that the parameters of the subsystem are adequately scaled. In particular, we found that there are two methods (the copy number model and the association rate model), which can be applied to adjust the parameters and that the copy number model can be used for highly abundant TFs (such as the non-cognate TFs in this case), while the association rate model for lower abundant TFs (lacI in this case). 

To simulate non-cognate crowding we considered the following abundances for these TFs: \one $0$, \two $10^{4}$, \three $3\times 10^{4}$, \four $5\times 10^{4}$ and \five $7\times 10^{4}$ molecules. The association rate was set to the values listed in Table \ref{tab:subSystemParams}. This abundance of non-cognate TFs, the corresponding association rates and the fact that each molecules covers $46\ bp$ of DNA lead to various percentages of DNA being covered, which reside in the range of biologically plausible values of $10\%$ to $50\%$ \citep{flyvbjerg_2006} (except in the case of $TF_{\textrm{nc}}=0$); see Table \ref{tab:subSystemParams}.

\begin{table*}
\begin{center}
  \begin{tabular}{ | r | r | r | r | r | r | r | r |}
\hline
$TF_{\textrm{nc}}$ & $k^{\textrm{assoc}}_{\textrm{nc}}\ s^{-1}$ & \textbf{covered DNA} & $\overline{TF}_{\textrm{nc}}$& $\overline{k}^{\textrm{assoc}}_{1\textrm{lacI}}\ s^{-1}$& $\overline{k}^{\textrm{assoc}}_{10\textrm{lacI}}\ s^{-1}$& $\overline{k}^{\textrm{assoc}}_{100\textrm{lacI}}\ s^{-1}$& $\overline{k}^{\textrm{assoc}}_{1000\textrm{lacI}}\ s^{-1}$\\ \hline
$0$ & $1800$ & $0\%$ & $0$ & $4.19$ & $4.04$ & $4.11$ & $4.19$\\ \hline
$10000$ & $2000$ & $9\%$ & $216$ & $4.58$ & $4.63$ & $4.67$ & $4.74$\\ \hline
$30000$ & $2571$ & $26\%$ & $647$ & $6.11$ & $6.10$ & $6.19$ & $6.32$\\ \hline
$50000$ & $3600$ & $42\%$ & $1078$ & $8.63$ & $8.76$ & $8.73$ & $8.88$\\ \hline
$70000$ & $6000$ & $55\%$ & $1509$ & $13.15$ & $13.05$ & $13.06$ & $13.26$\\ \hline
      \end{tabular}
\end{center}
\caption{\emph{Sub-system parameters for various non-cognate molecule abundances} The overbar is used to denote the corresponding parameters in the subsystem, e.g. $\overline{TF}_{\textrm{nc}}$ represents the abundance of non-cognate TFs in the $100\ Kbp$ subsystem. } \label{tab:subSystemParams}
\end{table*}

For each set of parameters, we performed $50$ simulations, each running for $3000\ s$, which is approximately the \emph{E.coli} cell cycle \citep{rosenfeld_2005}. To increase simulation speed, we selected a $100\ Kbp$ region of DNA which contained the $O_1$ site (nucleotides $300,000-400,000$ in the \emph{E.coli} K-12 genome) \citep{riley_2006}. Since the non-cognate TFs are highly abundant, we applied the copy number model and obtained the corresponding abundances of non-cognate TFs ($\overline{TF}_{\textrm{nc}}$) for use in the subsystem, as listed in Table \ref{tab:subSystemParams}.

In addition to non-cognate TFs, the system also consists of cognate lacI molecules. We considered several lacI abundances: \one $1$, \two $10$, \three $100$ and \four $1000$ molecules. As well as in the case of non-cognate TFs, we used the same parameters for lacI as in previous work \citep{zabet_2012_model,zabet_2012_subsystem}. In the case of the full system we considered an association rate of $k^{\textrm{assoc}}_{\textrm{lacI}}=2400\ s^{-1}$. When we applied the association rate model to reduce the system to $100\ Kbp$, we obtained the values of the association rate corresponding to each of the cases listed in Table \ref{tab:subSystemParams}. 

\subsection{Immobile obstacles}
We also considered the case of \emph{immobile} non-cognate molecules. These molecules  are bound to the DNA at a random position \citep{berg_1981} when the simulations start, and stay at that position until the simulations end. We allow immobile non-cognate TFs to cover the $O_1$ site and, due to the fact that the model implements steric hindrace, the binding of any immobile non-cognate molecule within $T_{\textrm{lacI}}^{\textrm{size}}+T_{\textrm{nc}}^{\textrm{size}}-1 = 66\ bp$ around the  $O_1$ site would exclude lacI molecules indefinitely from the $O_1$ site.

For immobile obstacles, we performed $1000$ simulations for each set of parameters and simulations where the target site is never reached are discarded.  We found that, in the most extreme cases, ($70000$ immobile non-cognate molecules) only $10\%$ of the simulations lead to the target site being occupied by lacI within $3000\ s$; see Figure \ref{fig:TS3Dcount}. Note that the number of simulations in the case of immobile obstacles is significantly higher compared to the mobile obstacles case. The main reason for that is that the simulation time is significantly shorter in the case of immobile obstacles compared to the case of mobile obstacles; i.e., in the case of immobile obstacles a simulation of $3000\ s$ takes in the orders of hours, while, in the case of mobile obstacles, a simulation takes in the order of several weeks.

In the case of immobile obstacles, we also consider the case of $40 000$ copies of non-cognate TFs. This was justified by the fact that, in the case of immobile obstacles, due to high residence time of the non-cognate TF molecules to the DNA, the percentage of DNA covered by molecules was higher than in the case of mobile obstacles. For $40000$ copies of non-cognate immobile molecules, $40\%$ of the DNA was covered by DNA binding molecules, which is similar to the crowding level observed in the case of $50000$ copies of mobile non-cognate molecules.  When we applied the copy number model and the association rate model to reduce the system to $100\ Kbp$, we obtained the following values: \one $\overline{TF}_{\textrm{nc}}=863$ and \two $\overline{k}^{\textrm{assoc}}_{1\textrm{lacI}}=\overline{k}^{\textrm{assoc}}_{10\textrm{lacI}}=\overline{k}^{\textrm{assoc}}_{100\textrm{lacI}}=\overline{k}^{\textrm{assoc}}_{1000\textrm{lacI}} = 7.37$. Note that in the case of immobile obstacles, the association rate affects the results negligibly as long as the binding to the DNA is fast compared to the amount of time spent bound to the DNA.

\section{Acknowledgments}
We would like to thank Mark Calleja for his support with configuring our simulations to run on CamGrid and Rob Foy, Robert Stojnic and Daphne Ezer for useful discussions and comments on the manuscript.

\emph{Funding:} This work was supported by Medical Research Council [G1002110 to N.R.Z.]. B.A. is a Royal Society University Research Fellow.

\clearpage

\appendix
    \begin{center}
      {\bf APPENDIX}
    \end{center}
\supportingsection

\section*{TF parameters} \label{sec:AppendixTFparams}

The default parameters used here were previously derived in \citep{zabet_2012_model} and \citep{zabet_2012_subsystem} and are listed in Table \ref{tab:modelTFparams}. In order to compare our results to the ones of \citet{li_2009} and \citet{murugan_2010}, we considered also a \emph{pseudo-immobile} non-cognate TF species. We assumed that this species does not diffuse on the DNA (hence the $P^{\textrm{left}}_{inc}=P^{\textrm{right}}_{inc}=0.0$, $P^{\textrm{unbind}}_{inc}=1.0$ and $P^{\textrm{jump}}_{inc}=1.0$). In order to increase the  the amount of time spent at one position, we $\tau^{0}_{inc}=13.2e+09$. We derived this waiting time by assuming that if the affinity landscape has a mean binding energy of $13\ K_BT$ and we aim to keep the molecules bound to the DNA $30000\ s$ (ten times the length of the simulation), then \citep{zabet_2012_model}
\begin{equation}
30000=\tau^{0}_{inc}\exp{(-13)}\Rightarrow \tau^{0}_{inc}=30000\cdot \exp{(13)} \approx 13.2e+09
\end{equation}

\begin{table}[h]
\centering
\begin{tabular}{| p{5cm} | r | r | r | l |}
\hline
\textbf{parameter} & \textbf{lacI} & \textbf{non-cognate}& \textbf{immobile non-cognate} &  \textbf{notation}\\ \hline
copy number &  $\in\{1,10,100,1000\}$ &   \multicolumn{2}{|c|}{see Table \ref{tab:subSystemParams}} & $TF_{x}$\\ \hline
motif sequence &  see Table \ref{tab:lacI3gPWM} & - & - &\\ \hline
energetic penalty for mismatch &  $1\ K_BT$ & $13\ K_BT$ & $13\ K_BT$ & $\varepsilon^{*}_{x}$\\ \hline
nucleotides covered on left &  $0\ bp$ & $23\ bp$ & $23\ bp$ & $TF^{\textrm{left}}_{x}$\\ \hline
nucleotides covered on right &  $0\ bp$ & $23\ bp$ & $23\ bp$ & $TF^{\textrm{right}}_{x}$\\ \hline
association rate to the DNA &    \multicolumn{3}{|c|}{see Table \ref{tab:subSystemParams}}  & $k^{\textrm{assoc}}_{x}$\\ \hline
unbinding probability &  $0.001474111$ & $0.001474111$ & $1.0$ & $P^{\textrm{unbind}}_{x}$\\ \hline
probability to slide left &  $0.4992629$ & $0.4992629$ & $0.0$ & $P^{\textrm{left}}_{x}$\\ \hline
probability to slide right &  $0.4992629$ & $0.4992629$ &  $0.0$ & $P^{\textrm{right}}_{x}$\\ \hline
probability to dissociate completely when unbinding &  $0.1675$ & $0.1675$ & $1.0$ & $P^{\textrm{jump}}_{x}$\\ \hline
time bound at the target site &  $1.18E-6\ s$ & $ 0.3314193\ s$ & $13.2e+09$ & $\tau^{0}_{x}$\\ \hline
the size of a step to left &  $1\ bp$ & $1\ bp$ & $1\ bp$ & \\ \hline
the size of a step to right &  $1\ bp$ & $1\ bp$ & $1\ bp$ & \\ \hline
variance of repositioning distance after a hop &  $1\ bp$ & $1\ bp$ & $1\ bp$ & $\sigma^{2}_{\textrm{hop}}$\\ \hline
the distance over which a hop becomes a jump  &  $100\ bp$ & $100\ bp$ & $100\ bp$ & $d_{\textrm{jump}}$\\  \hline
the proportion of prebound molecules & $0.0$ &  $0.9$ & $0.9$ &\\  \hline
affinity landscape roughness & - & $1.0\ K_BT$ & $1.0\ K_BT$ &\\ 
\hline
\end{tabular}
\caption{\emph{TF species default parameters}}
\label{tab:modelTFparams}
\end{table}


The PWM of the lacI was presented in \citep{zabet_2012_subsystem} and is also listed in Table \ref{tab:lacI3gPWM}.

\begin{table*}
\begin{center}
  \begin{tabular}{ |r | r | r | r | r |}
    \hline
     &   \multicolumn{4}{|c|}{\textbf{PWM}}  \\ \hline
    \textbf{Position} & A  &  C & G & T  \\ \hline
    $1$ & $0.6200$ & $-0.6900$ & $0.1400$ & $-0.6900$\\ \hline
    $2$ & $0.6200$ & $-0.6900$ & $0.1400$ & $-0.6900$\\ \hline
    $3$ & $0.1600$ & $0.1400$ & $-0.6900$ & $0.1800$\\ \hline
    $4$ & $0.1600$ & $-0.6900$ & $-0.6900$ & $0.6200$\\ \hline
    $5$ & $-0.7000$ & $-0.7000$ & $0.9000$ & $-0.7000$\\ \hline
    $6$ & $-0.6900$ & $-0.6900$ & $-0.6900$ & $0.9300$\\ \hline
    $7$ & $0.0077$ & $-0.0084$ & $-0.0073$ & $0.0083$\\ \hline
    $8$ & $0.0077$ & $-0.0084$ & $-0.0073$ & $0.0083$\\ \hline
    $9$  & $0.0077$ & $-0.0084$ & $-0.0073$ & $0.0083$\\ \hline
    $10$ & $0.0077$ & $-0.0084$ & $-0.0073$ & $0.0083$\\ \hline
    $11$ & $0.0077$ & $-0.0084$ & $-0.0073$ & $0.0083$\\ \hline
    $12$ & $0.0077$ & $-0.0084$ & $-0.0073$ & $0.0083$\\ \hline
    $13$ & $0.0077$ & $-0.0084$ & $-0.0073$ & $0.0083$\\ \hline
    $14$ & $0.0077$ & $-0.0084$ & $-0.0073$ & $0.0083$\\ \hline
    $15$ & $0.0077$ & $-0.0084$ & $-0.0073$ & $0.0083$\\ \hline
    $16$ & $0.6200$ & $-0.6900$ & $0.1400$ & $-0.6900$\\ \hline
    $17$ & $-0.7000$ & $0.9000$ & $-0.7000$ & $-0.7000$\\ \hline
    $18$ & $0.9300$ & $-0.6900$ & $-0.6900$ & $-0.6900$\\ \hline
    $19$ & $0.9300$ & $-0.6900$ & $-0.6900$ & $-0.6900$\\ \hline
    $20$ & $-0.6900$ & $0.1400$ & $-0.6900$ & $0.6200$\\ \hline
    $21$ & $-0.6900$ & $0.1400$ & $-0.6900$ & $0.6200$\\ \hline
  \end{tabular}
\end{center}
\caption{lacI PWM} \label{tab:lacI3gPWM}
\end{table*}

\section*{The number of simulations where the target site was reached}  \label{sec:AppendixTFimobileReached}

\begin{figure}
\centering
\includegraphics[angle=270,width=\textwidth]{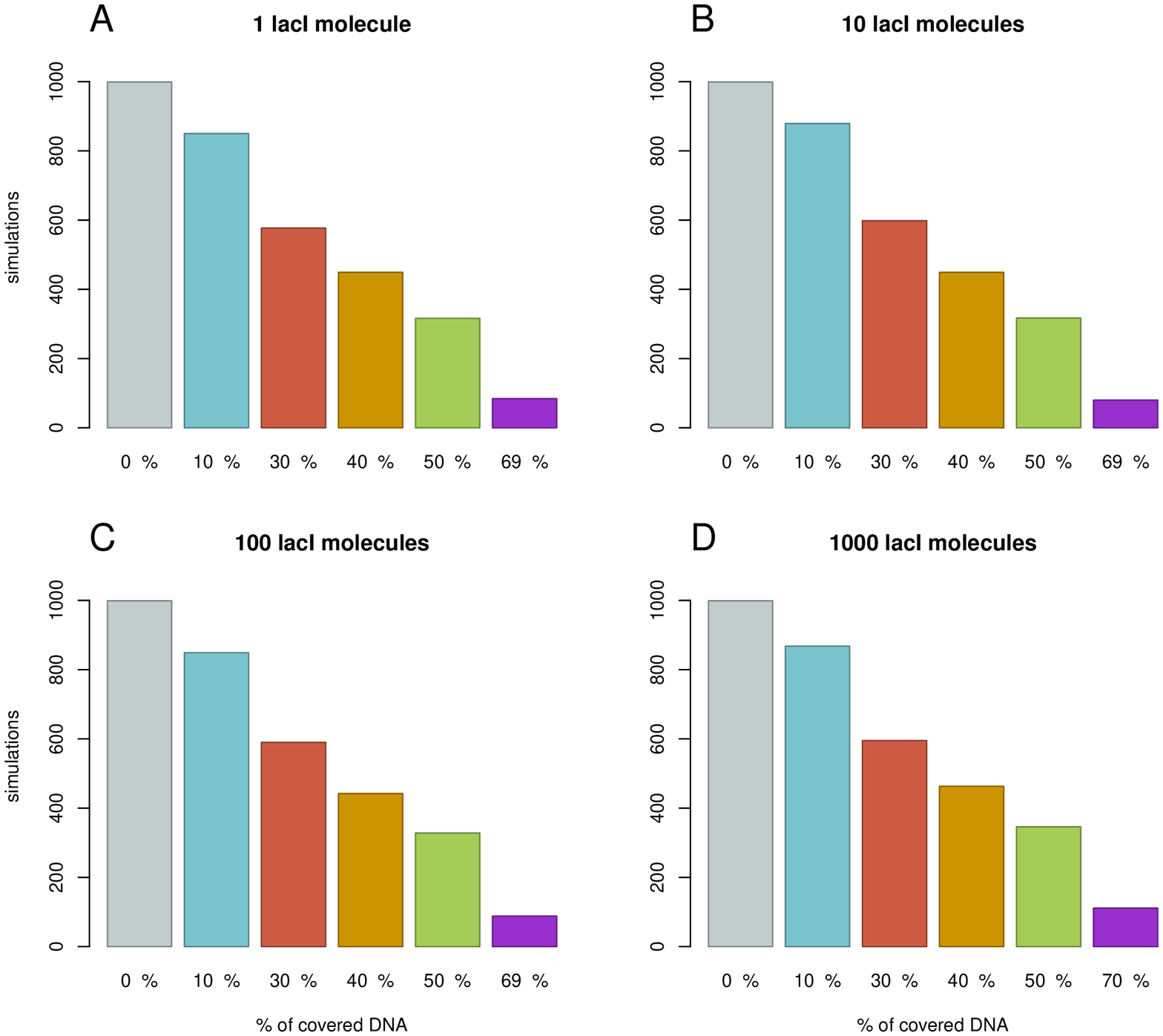}
\caption{\emph{The number of simulations where the target site was reached within $3000\ s$ in the case of immobile obstacles}.  \label{fig:TS3Dcount}}
\end{figure}

When a molecule binds to the DNA, it is uniformly distributed between all available position \citep{berg_1981,zabet_2012_model}. This means that in the case of immobile obstacles, when the non-cognate molecules get bound to the DNA, there is a probability that they will bind to the target site ($O_1$) and, thus, in those simulations the target site is unreachable. We removed these points from the data and found that by increasing the number of non-cognate molecules, the probability of covering the target site also increases. Figure \ref{fig:TS3Dcount} confirms that, by increasing the crowding on the DNA, the number of simulations that resulted in the binding of lacI to $O_1$ site within $3000\ s$ decreases.

We found that the proportion of simulations that resulted in the location of the target site within $3000\ s$ is approximately: \one $p_{reached}^{0.1}=0.85$, \two $p_{reached}^{0.3}=0.59$, \three $p_{reached}^{0.4}=0.44$, \four $p_{reached}^{0.5}=0.33$ and \five $p_{reached}^{0.7}=0.1$ (where the superscript indicates the proportion of DNA that is covered by DNA binding molecules); see Figure \ref{fig:TS3Dcount}. 

\section*{Search time in the case of immobile obstacles}  \label{sec:AppendixTFimmobileTime}

\begin{figure}
\centering
\includegraphics[angle=270,width=\textwidth]{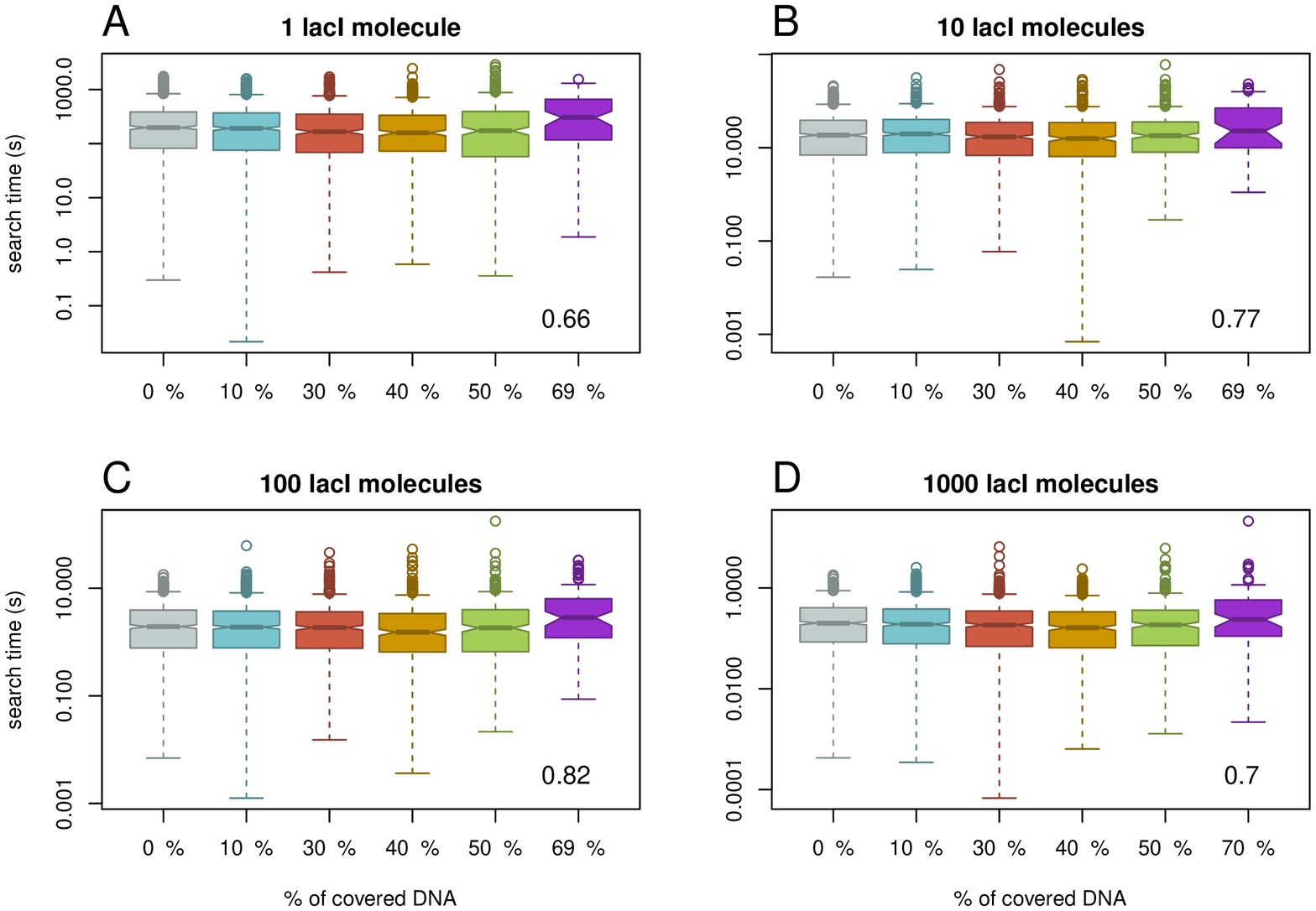}
\caption{\emph{The average time for the TF to reach the target site (measured in seconds) as a function of DNA crowding in the case of immobile obstacles}. For each set of parameters, we performed $1000$ simulations. Note that the amount of covered DNA is higher than in the case of mobile obstacles, due to the fact that the molecules spend more time bound to the DNA. It should be noted that in the case of $70\%$ of the DNA being covered by DNA binding proteins, the probability to locate the target site within a cell cycle is as low as $0.1$; see Figure \ref{fig:TS3Dcount}. The number in the inset represents the Pearson coefficient of correlation between crowding and the mean of the search time.  Note that to enhance the visibility, the boxplots are positioned equidistant although the crowding levels are not. \label{fig:TStimeImmobile}}
\end{figure}

Figure \ref{fig:TStimeImmobile} shows the search time as a function of crowding levels on the DNA in the case of immobile obstacles. 

\section*{Statistical significance of the change in the search time}  \label{sec:AppendixANOVAtime}

\begin{figure}
\centering
\includegraphics[angle=270,width=0.5\textwidth]{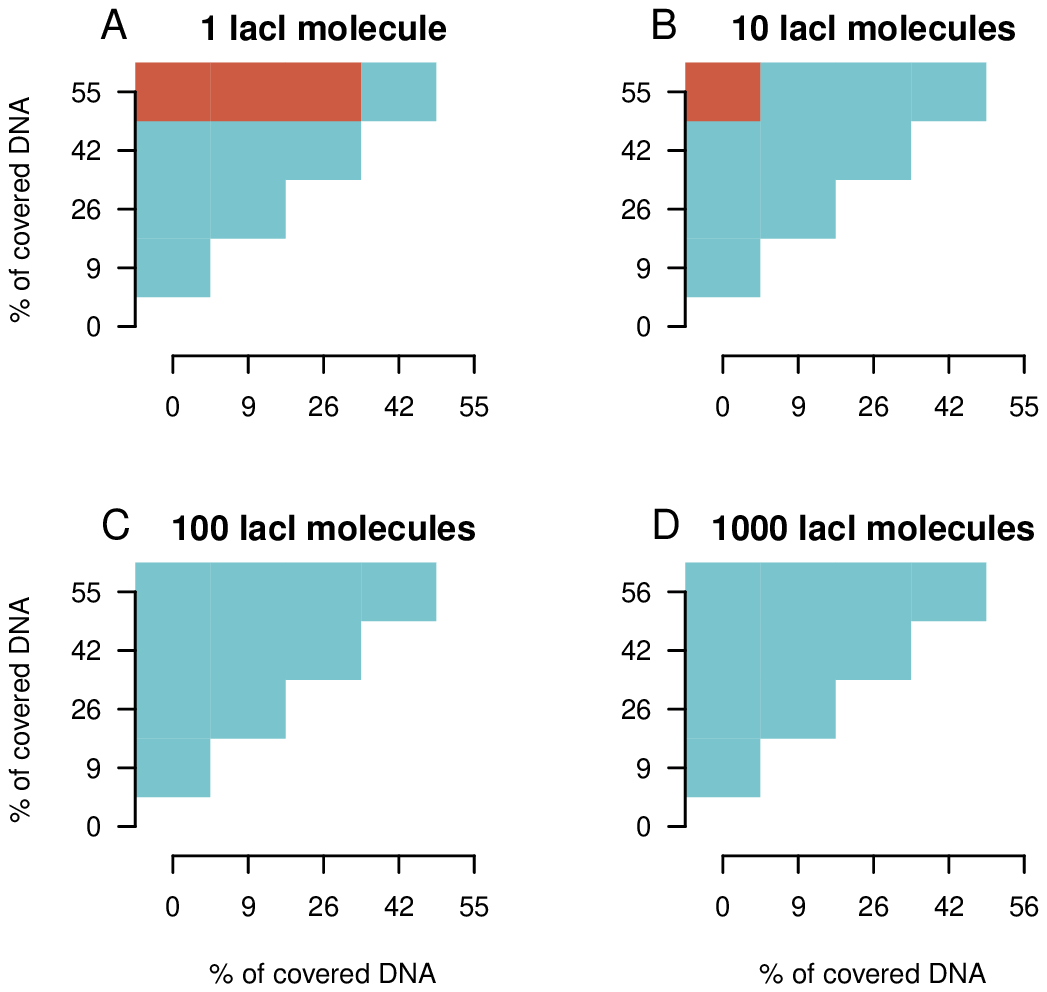}
\caption{\emph{Statistical significance of the change in the search time in the case of mobile obstacles}. The graph represents the pairwise statistical test between the distributions of arrival times to the target site at various crowding levels.  We performed Tukey's range test (for a $95\%$ confidence interval) on a one-way ANOVA of the logarithm of the search time. The color indicates the p-value of the difference between the corresponding search times. We represent by red the case of p-values lower than $0.05$, and by blue the case of  p-values higher than $0.05$. The graph confirms that for crowding levels on the DNA between $10\%$ and $50\%$ there is no statistically significant difference in the arrival times to the target site. \label{fig:TStimeANOVA}}
\end{figure}

\begin{figure}
\centering
\includegraphics[angle=270,width=0.5\textwidth]{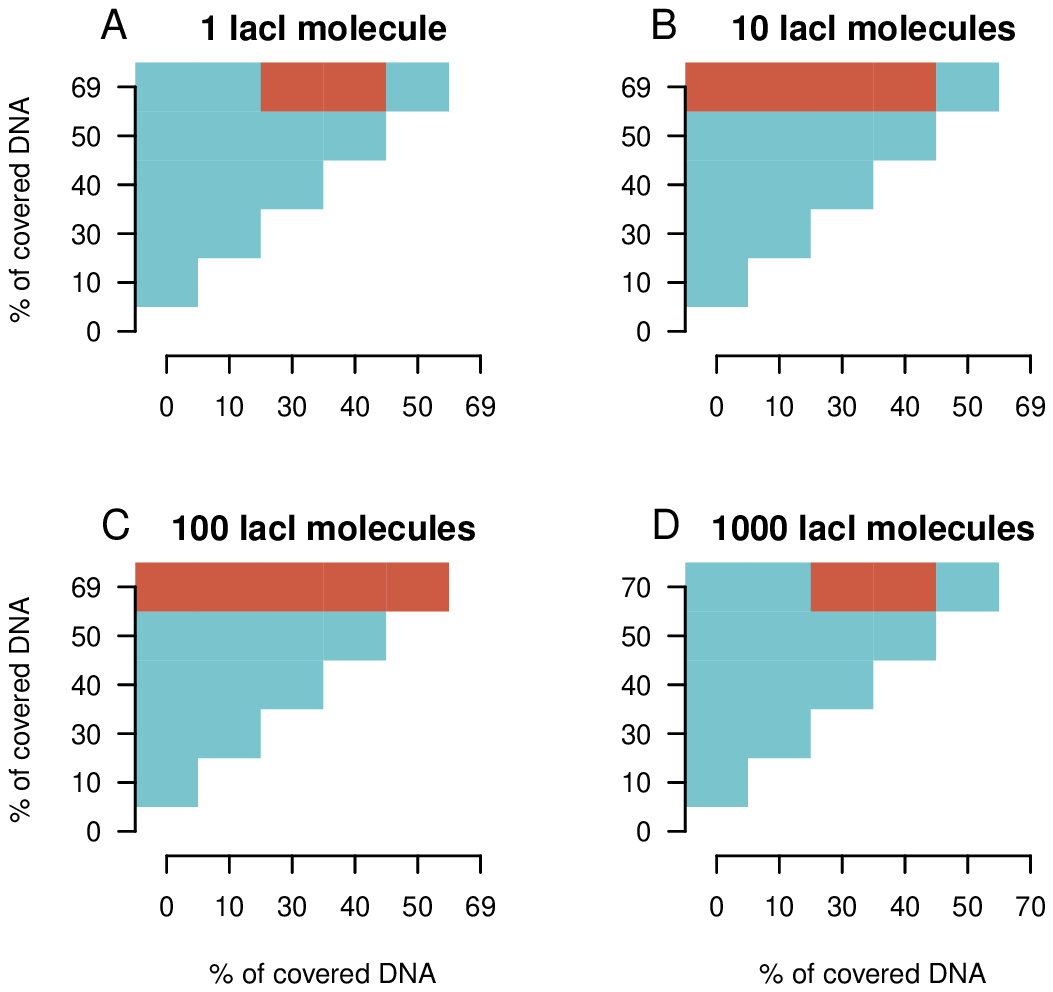}
\caption{\emph{Statistical significance of the change in the search time in the case of immobile obstacles}. The graph represents the pairwise statistical test between the distributions of arrival times to the target site at various crowding levels.  We performed the Tukey's range test on the logarithm of the search time and the color indicates the p-value of the difference between the corresponding search times. We represent by red the case of p-values lower than $0.05$ and by blue the case of  p-values higher than $0.05$. The graph confirms that for crowding levels on the DNA between $10\%$ and $50\%$ there is no statistically significant difference in the arrival times to the target site. \label{fig:TStimeANOVA3D}}
\end{figure}

Figure \ref{fig:TStimeANOVA} and Figure \ref{fig:TStimeANOVA3D} confirm that for crowding levels on the DNA between $10\%$ and $50\%$ there is no statistically significant difference in the search time. 

\section*{Statistical significance of the change in the occupancy of the target site}  \label{sec:AppendixANOVAoccupancy}

\begin{figure}
\centering
\includegraphics[angle=270,width=0.5\textwidth]{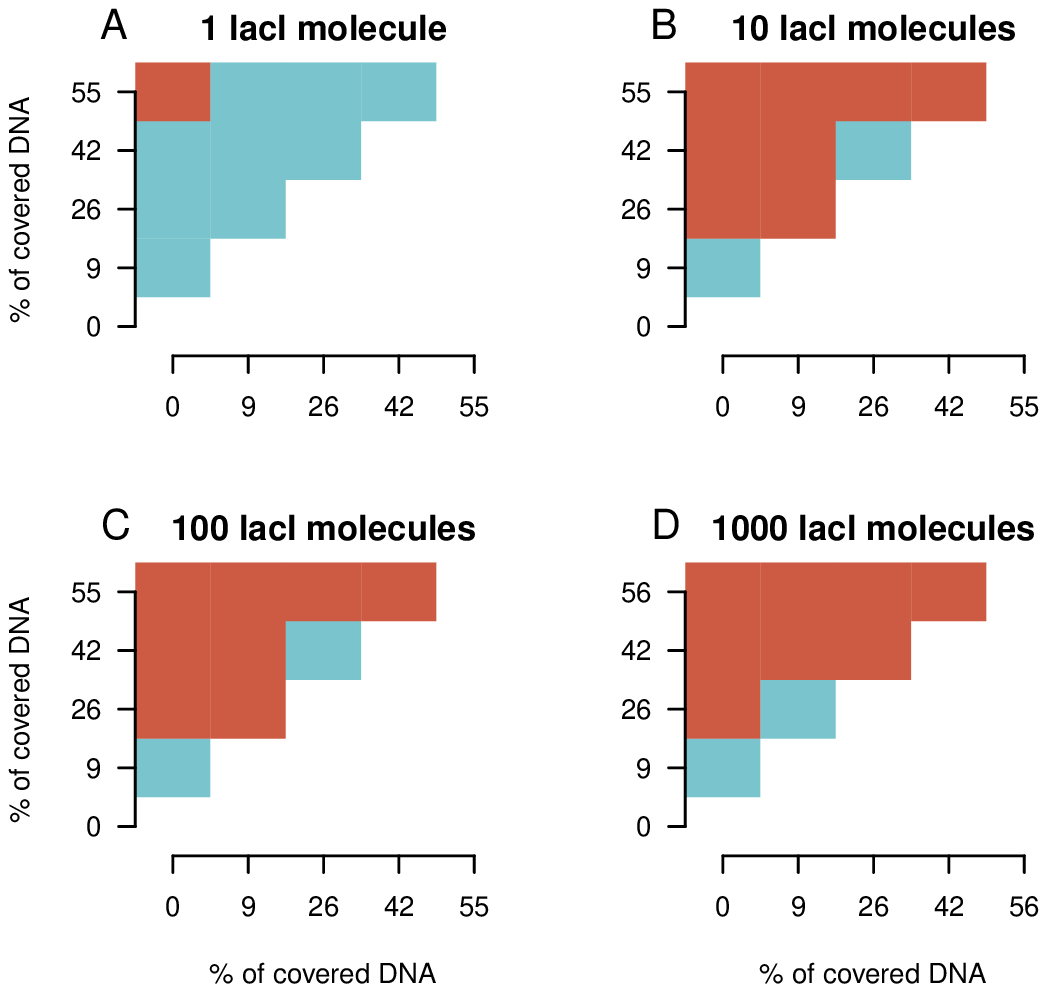}
\caption{\emph{Statistical significance of the change in the proportion of time the target site is occupied in the case of mobile obstacles}. The graph represents the pairwise statistical test between the distributions of occupancy of the target site at various crowding levels. We performed the Tukey's range test on the logarithm of the occupancy of the target site and the color indicates the p-value of the difference between the corresponding occupancies of the target sites. We represent by red the case of p-values lower than $0.05$ and by blue the case of  p-values higher than $0.05$. The graph confirms that the crowding levels considered lead to statistically significant difference in the occupancy of the target site, except for the case of $1$ lacI molecules (usually associated with leaky expression of the gene encoding the TF). \label{fig:TSproportionANOVA}}
\end{figure}

\begin{figure}
\centering
\includegraphics[angle=270,width=0.5\textwidth]{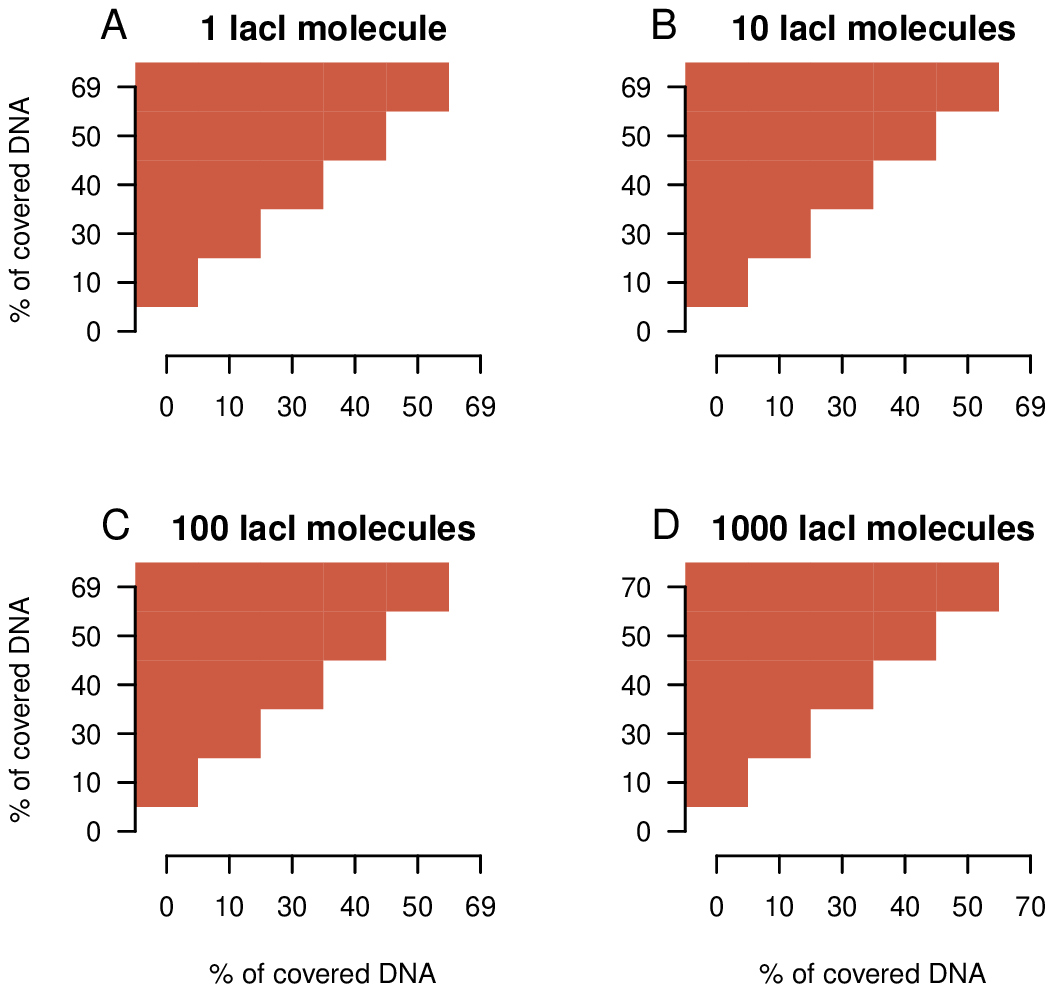}
\caption{\emph{Statistical significance of the change in the proportion of time the target site is occupied in the case of immobile obstacles}. The graph represents the pairwise statistical test between the distributions of occupancy of the target site at various crowding levels.  We performed the Tukey's range test on the logarithm of the occupancy of the target site and the color indicates the p-value of the difference between the corresponding  occupancies of the target sites. We represent by red the case of p-values lower than $0.05$ and by blue the case of  p-values higher than $0.05$. The graph confirms that the crowding levels considered lead to statistically significant difference in the occupancy of the target site. \label{fig:TSproportionANOVA3D}}
\end{figure}

Figure \ref{fig:TSproportionANOVA} and Figure \ref{fig:TSproportionANOVA3D} confirm that for  biologically relevant  crowding levels on the DNA, there is a statistically significant difference in the occupancy of the target site. Note that for 1 molecule of lacI and mobile obstacles, the crowding level does not significantly change the occupancy of the target site; see Figure \ref{fig:TSproportionANOVA}\onel. 

\section*{Comparison between the mobile and immobile obstacle case}  \label{sec:AppendixImmobileMobile}

\begin{figure}
\centering
\includegraphics[angle=270,width=\textwidth]{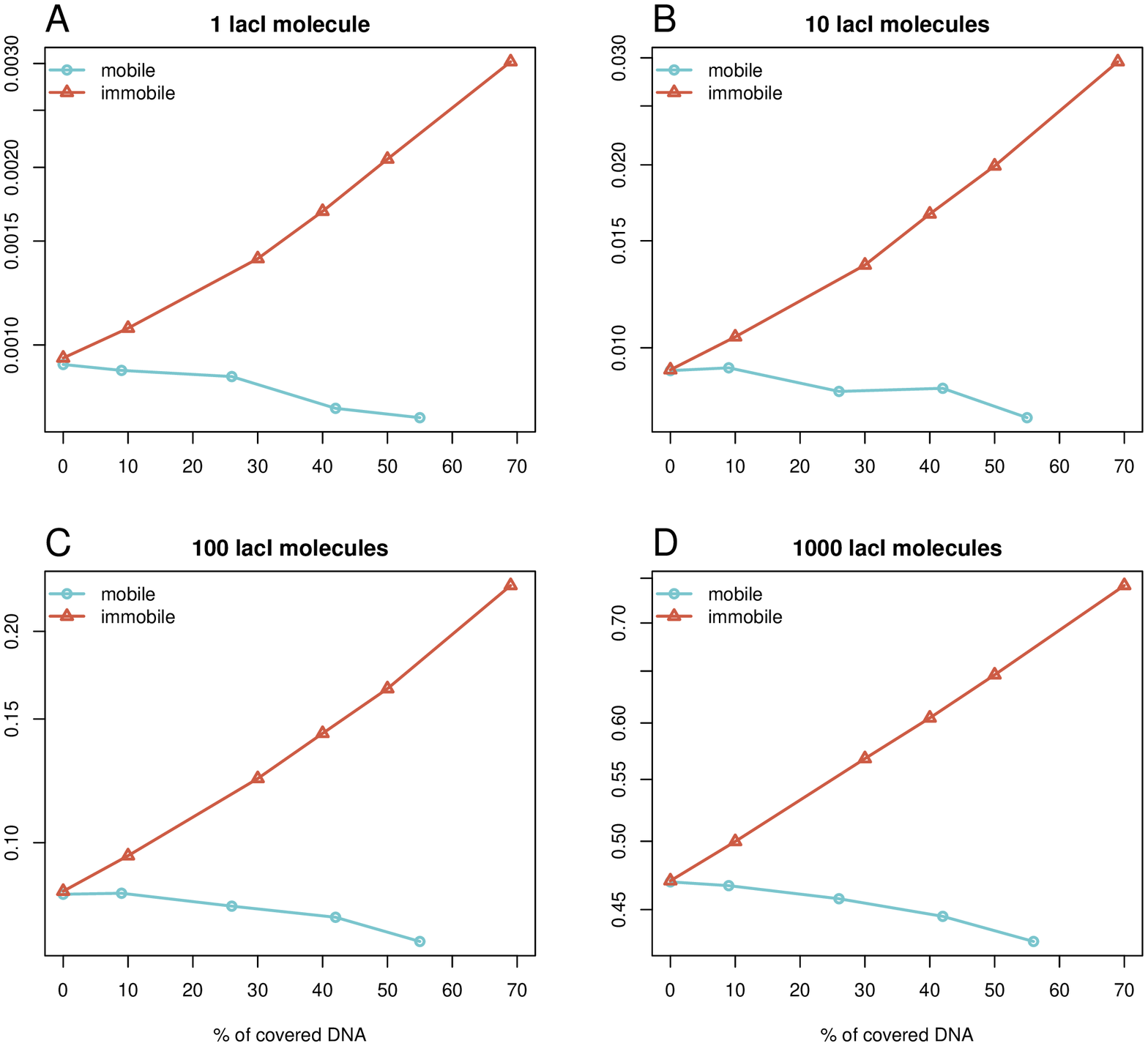}
\caption{\emph{The comparison between the mean occupancy time of the target site in the case of mobile obstacles and in the case of immobile obstacles}.  \label{fig:TSproportionComparison}}
\end{figure}

Finally, we compared the overall mean occupancy of the target site between the case of mobile and immobile obstacles. Our results showed that,  when the obstacles are fixed on the DNA, the occupancy of the target site is higher (see Figure \ref{fig:TSproportionComparison}).

\section*{Proportion of time the target site is occupied in the case of mobile obstacles}  \label{sec:AppendixOccupancyNoise}

\begin{figure}
\centering
\includegraphics[angle=270,width=\textwidth]{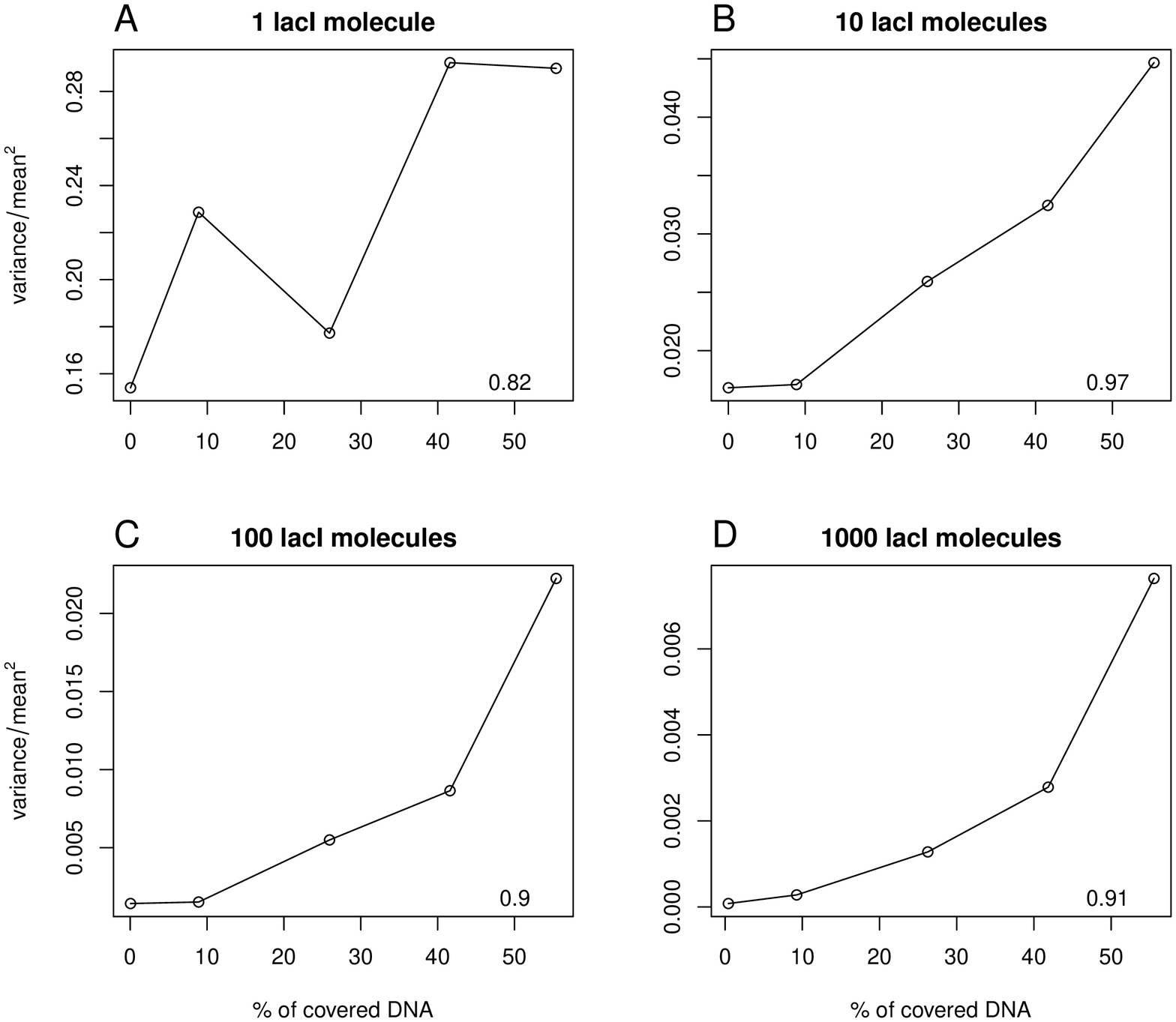}
\caption{\emph{Noise in the proportion of time the target site is occupied as a function of the crowding levels on the DNA in the case of mobile obstacles}.   We normalised the variance by the square of the mean  as proposed in \citep{paulsson_2005}. The number in the inset represents the Pearson coefficient of correlation between crowding and the noise in the proportion of time the target site is occupied. \label{fig:TSproportionNoise}}
\end{figure}

\begin{figure}
\centering
\includegraphics[angle=270,width=\textwidth]{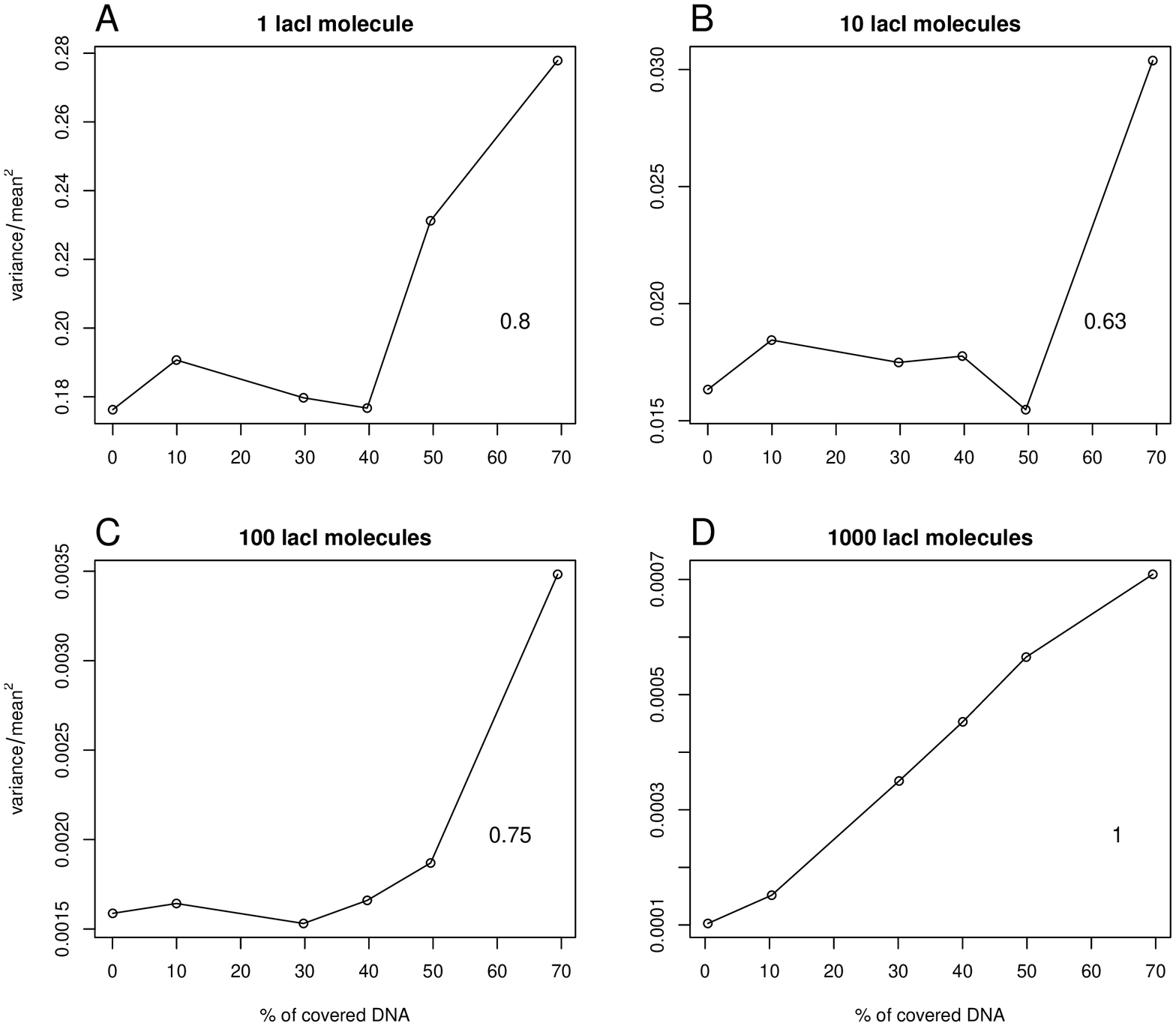}
\caption{\emph{Noise in the proportion of time the target site is occupied as a function of the crowding levels on the DNA in the case of immobile obstacles}.   We normalised the variance by the square of the mean. The number in the inset represents the Pearson coefficient of correlation between crowding and the noise in the proportion of time the target site is occupied. \label{fig:TSproportionNoise3D}}
\end{figure}

We also looked at the noise in occupancy and found that indeed, there is a strong correlation between crowding levels on the DNA and noise in the proportion of time the target site is occupied. In particular, we found that by increasing the level of crowding on the DNA the noise in the occupancy of the target site is increased; see Figure \ref{fig:TSproportionNoise}. Interestingly, this is valid for both mobile (Figure \ref{fig:TSproportionNoise}) and immobile obstacles (Figure \ref{fig:TSproportionNoise3D}). 

\bibliographystyle{apalike}
\bibliography{grip_crowding.bib}

\end{document}